\magnification=1200
\overfullrule0pt
\baselineskip18truept
\pageno=4
\hsize16truecm
\vsize23truecm

\font\bb=msbm10
\def\bs{\vskip24pt}

\def\ss{\vskip6pt}
\def\nt{\noindent}

\def\Z{\hbox{\bb Z}}

\def\cl{\centerline}

\def\b#1{\hbox to 0pt{$#1$\hss}\mkern.65mu\hbox to 0pt{$#1$\hss}\mkern.65mu#1}
\topinsert
\vskip.3in
\endinsert
\cl{\bf THE SEMISPIN GROUPS IN STRING THEORY} 
\vskip0.3in
\cl{\bf Brett McInnes \footnote{$^*$}{matmcinn@nus.edu.sg} } 
\cl{\bf Department of Mathematics }
\cl{\bf National University of Singapore}
\cl{\bf 10 Kent Ridge Crescent, Singapore 119260} 
\cl{\bf Republic of Singapore}. 
\vskip0.4in
\bs
ABSTRACT  : \ \ In string theory, an important role is played by certain Lie groups which are locally isomorphic to $SO (4m), \ m \leq 8 $. It has long been known that these groups are actually isomorphic not to $SO (4m)$ but rather to the groups for which the half - spin representations are faithful, which we propose to call $Semispin(4m)$.(They are known in the physics literature by the ambiguous name of $`` Spin (4m) / \Z_2 " .)$ Recent work on string duality has shown that the distinction between $SO (4m)$ and $Semispin(4m)$ can have a definite physical significance. 
This work is a survey of the relevant properties of $Semispin(4m)$ and its subgroups. \hfill\break
PACS : 11.25.\ Sq, 02.\ 40.\ Ma

\vfill\eject

\nt{\bf I.\ \ INTRODUCTION.}
\bs
From a physical point of view, gauge theories have a serious drawback : the construction works equally well for all compact Lie groups. Even when the Lie algebra of the gauge group is known, the global structure of the group itself is not fixed by any fundamental principle. For example, it can be argued that the gauge group of the standard model [1] is ``really" $[ SU(3) \times  SU(2) \times  U(1) ] / \Z_6 $ rather than  $SU(3) \times SU(2) \times  U(1)$; but since all known particles fall into multiplets which can be regarded as representations of {\it both} of these groups, there is ( at present ) no way of deciding the issue other than by an appeal to parsimony.   
\ss
It is one of the many virtues of string theory that it puts an end to all uncertainty on this score [2]. The theory not only specifies the dimension of the gauge group (at 496 in the Type I and heterotic theories) but also its global structure within each version. In the heterotic $``E_{8} \times  E_{8}$" theory, this is quite straightforward. First, there is in any case {\it only one} connected group with the Lie algebra of $E_{8} \times E_{8}$, namely $E_8 \times E_8$ itself. Second, there is {\it only one} non-trivial disconnected Lie group with  $E_8 \times E_8$ as identity component, namely the semi-direct product $(E_8 \times  E_8) \triangleleft \Z_{2}$, where $\Z_{2}$ acts by exchanging the $E_8$  factors. As the corresponding string theory ( initially ) treats the two factors symmetrically, we conclude that the global version of the gauge group is $(E_{8} \times  E_{8})\triangleleft \Z_{2}$. ( The significance of disconnected gauge groups is discussed in Refs. [3],[4],[5],[6]. By ``non-trivial" we mean to exclude, for example, direct products of finite groups with  $E_{8} \times  E_{8}$, which are of little or no physical interest.) 
\ss
The $``SO(32)"$ cases ( Type I and heterotic ) are much less straightforward, because there are many non-trivial groups with the same Lie algebra as $SO(n)$; in fact, there are eight non-trivial groups with the $``SO(32)"$ algebra. It has been known from the beginning [7] that string theory selects from these eight a group known in the physics literature as $Spin (32)/ \Z_2$ - a most unfortunate convention, which only exacerbates the tendency to confuse this group with  $SO(32)$. This group is closely associated with ( and is essentially defined by ) the {\it half-spin} representations of $Spin(32)$, and so we propose the name $Semispin(32)$ for it. As we shall see, there is {\it no} non-trivial disconnected Lie group with $Semispin (32)$ as identity component, so the gauge group is connected in this case.
\ss
To summarise : the heterotic string theories fix the global structures of their gauge groups. One theory uses the disconnected but simply connected group  $(E_{8} \times E_{8})\triangleleft \Z_2$, while the other uses the connected but not simply connected group $Semispin (32)$. 
\ss
The semispin groups are perhaps the least familiar of the compact simple Lie groups, and there is a venerable tradition of treating $Semispin(4m)$ as if it were the same as $SO(4m)$. We wish to argue that this tradition has outlived its usefulness, that string theory forces us to be fully aware of the differences between $Semispin(4m)$ and the other groups with the same algebra. There are two physically significant kinds of distinction, one representation - theoretic, the other topological.
\ss
First, note that while such ambiguities have often arisen in the past, one of the groups in question has always been a {\it cover} of the other. For example, $``SO(10)"$ grand unification [8] uses a certain 16 - dimensional multiplet which does {\it not} correspond to any representation of $SO(10)$. It is, of course, a representation of $Spin(10)$. One can solve this ``problem" by simply reading $Spin(10)$ for $SO(10)$; no harm is done, but only because every representation of 
$SO(10)$ is automatically a representation of $Spin(10)$, the latter being a {\it cover} of the former. In the opposite direction, one normally writes $SU(3) \times SU(2) \times  U(1)$ for $(SU(3) \times  SU(2) \times  U(1)/ \Z_6$, the ``true" group [1], with no ill effects because every representation of the latter is a representation of the former. The novelty in string theory is that {\it neither } $SO(32)$  {\it nor} $Semispin(32)$ is a cover of the other. Consequently, {\it both}  have representations which cannot be regarded as representations of the other. The situation here is quite different to the superficially analogous ambiguities arising in earlier gauge theories.
\ss
Secondly, there exist $Semispin (32)$ gauge configurations (over topologically non - trivial space - times ) which are of considerable physical importance, but which {\it cannot} be interpreted as $SO(32)$ gauge configurations [9],[10]. The reader might argue that one can likewise construct $SO(10)$ configurations which do not lift to $Spin(10)$. The point, however, is just that ordinary gauge theory does not provide any fundamental justification for thinking $SO(10)$ important. String theory, by contrast, does favour $Semispin(32)$ over $SO(32)$. The analysis of $Semispin(32)$ bundles which cannot be regarded as $Spin(32)$ or $SO(32)$ bundles is therefore physically significant. 
\ss
Finally, the study of duality [11] brings both points together in a potentially very confusing way. The T-duality between the two heterotic theories relies on relating $``E_{8} \times E_{8}$" and $``SO(32)"$ through their supposed common subgroup, $``SO(16) \times  SO(16)"$. A global investigation shows that {\it no such common subgroup exists}; worse still, neither of the actual respective subgroups covers the other ; worse yet again, each has representations which are {\it not} representations of the other, but which {\it are} crucial in establishing duality. Solving this problem leads to further topological obstructions, and, in the background, one has ``Wilson loops" behaving in a way that depends very delicately on the global structure of various  subgroups of $Semispin(32)$ and $Semispin(16)$. In short, the local simplicity of the duality argument conceals considerable complexity at the global level.
\ss
The purpose of this work is not to solve all of these problems, but rather to give a useful survey of those aspects of the Semispin  groups  ( and their subgroups ) which are most directly relevant to string theory. The main emphasis is on the structure of the groups themselves rather than their representations, since the latter are well understood and since it is the former which is needed for dealing with topological obstructions and for analysing the effect of Wilson loops. 
\ss
We begin with a brief survey of the family of non - trivial Lie groups with the algebra of $SO(n)$.
\bs
\nt{\bf II.\ \ GROUPS WITH THE ALGEBRA OF ${\bf SO(n)}$}.
\ss
In order to understand the ways in which $Semispin(4m)$ differs from the other groups with the same Lie algebra, it is useful to begin with a complete classification. We refer the reader to Ref.[12] for the basic techniques, or to Ref.[13] for a much simpler account.
\ss
We shall not assume that the gauge group is connected : we have already seen that this would not be justifiable in one heterotic theory. On the other hand, it is true that most disconnected Lie groups are of little physical interest. Every compact Lie group can be expressed as a finite union of connected components,
$$G \ = G_{0} \ \bigcup \ \gamma_{1} \bullet G_{0} \ \bigcup \ \gamma_{2} \bullet \ G_{0} \ 
\bigcup \ldots$$ 
\nt
where $G_{0}$ contains the identity and the $\gamma_{i}$ are not elements of $G_{0}$. The non-identity components of a gauge group are particularly important if space-time is not simply connected, since in that case parallel transport of particles around non-contractible paths ( ``Wilson loops" ) can affect conserved charges [3],[4],[5]. However, a given component,  $\gamma_{i} \bullet  G_{0}$, can only give rise to such effects if $\gamma_{i}$ ${\it cannot}$ be chosen so as to commute with every element of $G_{0}$. The physically interesting disconnected groups are those such that none of the $\gamma_{i}$ can be chosen to commute with every element of $G_{0}$. Such a group is called a ${\it natural}$ ${\it extension}$ of its identity component. For example, $(E_8 \times E_8) \triangleleft \Z_{2}$ is a natural extension of $E_{8} \times E_{8}$, and it is in fact the only other natural extension. ( It is convenient to adopt the convention that a connected group is a natural extension of itself. ) Henceforth, we confine attention to disconnected groups which are natural extensions of their identity components.
\ss
Next, some definitions. Let $Pin(n), n \geq 2$, be defined as usual [14] in terms of a Clifford algebra with a basis $ \{ {e_{i}} \}$. We can write $Pin(n)$ as a natural extension of $Spin (n)$, when $n$ is even :
$$Pin (2m) = Spin (2m) \bigcup \ e_{1} \bullet Spin (2m).$$
\nt
Notice that this is not necessarily a semi-direct product, since $(e_{1})^{2} = -1 \in Spin (2m)$. However, $Spin \ (2m) \triangleleft \Z_{2}$ can be defined (with the generator of $\Z_{2}$ acting in the same way as conjugation by $e_{1}$); it is actually isomorphic to $Pin(2m)$ if $m$ is even, but not if $m$ is odd. It, too, is a natural extension of  $Spin(2m)$. ( There is no natural extension of $Spin(n)$, other than itself, when $n$ is odd. )  
\ss
Let ${\hat K}_{m,n}$ be defined by 
$${{\hat K}_{m,n} = {\mathop{\Pi}\limits_{i=m}^{n}} \ e_i}$$ 
\nt 
and set ${\hat K}_{m} = {\hat K}_{1,m}$. Then the centre of $Spin (n)$ is $ \{ \pm 1 \}$ if $n$ is odd, while the centre of $Spin (2m)$ is $ \{ \pm 1, \pm \hat K_{2m} \}$. Since $({\hat K_{2m})^{2}}$  = $({-1})^{m}$, the centre is $\Z_4$ if $m$ is odd, but $\Z_{2} \times \Z_{2}$ if $m$ is even [15]. Here we think of $\{ 1, \hat  K_{2m} \}$ as the first $ \Z_{2}$, $\{ 1, - \hat K_{2m} \}$ as the second, and $ \{ \pm 1 \}$ as the diagonal. Of course, we have 

$$Spin(n)/ \{ \pm 1 \} = SO(n) \ {\rm for \ all}\ n \geq 2.$$
\nt
When $n$ is odd, $SO(n)$ has no natural extension other than itself, but when $n$ is even it has two others. The first is $O(2m)$, which may be expressed as \hfill\break
$${ O(2m) = SO(2m) \bigcup A_{2m} \bullet SO(2m)},$$
\nt
where ${A_{2m}}$ is a ${(2m) \times (2m)}$ orthogonal matrix satisfying ${A_{2m}^{2}= I_{2m}, det A_{2m} = -1}$. Thinking of ${O(2m)}$ as the real subgroup of ${U(2m)}$, we can also define  

$${ Oi(2m) = SO(2m) \bigcup iA_{2m} \bullet SO(2m)};$$
\nt
this group is also a natural extension of $SO(2m)$, and it is not isomorphic to $O(2m)$.
\ss
When $n$ is even, $SO(n)$ has a non-trivial quotient, 

$$PSO(2m) = SO(2m)/ \{ \pm I_{2m} \},$$
\nt
the projective special orthogonal group. We can define $PO(2m)$ as the same quotient of $O(2m)$, and it is a natural extension of $PSO(2m)$. Notice that $PSO(2m)$ can be obtained directly from $Spin(2m)$ by factoring out the entire centre. 
\ss
When $n$ is a multiple of $4$, we can also consider the quotients $Spin(4m)/ \{ 1, \hat K_{4m} \}$ and  $Spin(4m)/ \{ 1, - \hat K_{4m} \}$. Let $Ad (e_1)$ denote conjugation by $e_{1}$ in $Pin(4m);$ then $Ad (e_1)$ is an automorphism of $Spin(4m)$, and 

$$Ad(e_1) \hat K_{4m} = - \hat K_{4m}.$$ 
\nt
It follows that $Spin (4m) / \{ 1, \hat K_{4m}\}$ and $Spin(4m)/ \{ 1, - \hat K_{4m}\}$ are mutually isomorphic. Thus we obtain only one group in this way, not two. We define

$$Semispin(4m) = Spin(4m) / \{ 1, \hat K_{4m}\}.$$
\nt 
This group is isomorphic to $SO(4m)$ only if $Spin(4m)$ admits an automorphism which maps $\hat K_{4m}$ to $-1$ ; but no such automorphism exists, except when $m=2$. Leaving that case to one side, $Ad(e_1)$ is, up to inner automorphisms, the only outer automorphism of $Spin(4m)$. Since $Ad(e_1)$ does not map $\{ 1, \hat K_{4m}\}$ into itself, we see that, unlike $Spin(4m), SO(4m)$, and $PSO(4m), Semispin(4m)$ has no outer automorphism if $m \not= 2$. If, therefore, $G$ is a compact disconnected group with $Semispin(4m)$ as identity component, 

$$G= Semispin(4m) \bigcup \gamma_{1} \bullet Semispin(4m) \bigcup \ldots,$$
\nt
then $Ad( \gamma_i)$ must, for all $i$, be inner : $Ad ( \gamma_i)= Ad (s_{i})$ for some $s_i$ in $Semispin(4m)$. Thus $\gamma_{i} \ s_{i}^{-1}$ commutes with every element of $Semispin(4m)$,  and so we see that, when $m \not= 2, Semispin(4m)$ has no natural extension other than itself.
\ss
When $m = 2$, we have $Spin(8)$, which has the triality map [14], an automorphism of order three. This combines with $Ad(e_{1})$ to give $D_{6}$, the dihedral group of order six. Triality  maps  $\hat K_8$ to $-1$, so in fact 
$$Semispin(8) = SO(8).$$
\nt
This is the only dimension in which the $Semispin$ construction gives nothing new. Triality does not descend to $SO(8)$ (because it does not preserve $\{ \pm 1 \})$ but it does descend to $PSO(8)$. 
\ss
We are now in a position to state the following theorem, the proof of which is an application of techniques given in Refs. [12] and [13]. 
\ss
\nt
THEOREM 1 : Let $G$ be a compact Lie group which is a natural extension of its identity        
            component. If the Lie algebra of $G$ is isomorphic to that of $SO(n)$, $n \geq 2$,             then $G$ is globally isomorphic to a group in the following list:
\ss
\nt
$(1) \ n=2 : \ SO(2), \ O(2), \ Pin(2)$. 
\ss
\nt
$(2) \ n=odd : \ SO(n), Spin(n)$.
\ss
\nt
$(3) \ n=4m+2, m \geq 1: \ Spin(n), \ Pin(n), \ Spin(n) \triangleleft \Z_{2}, \ SO(n), \ O(n), \ Oi(n),  \hfill\break PSO(n), \ PO(n)$.
\ss
\nt
$(4) \ n=4m, \ m \not= 2: \ Spin(n), \ Pin(n), \ SO(n), \ O(n), \ Oi(n), \ PSO(n),\  PO(n), \hfill\break
Semispin(n)$. 
\ss
\nt
$(5) \ n=8 : \ Spin(8), \  Pin(8), \ Spin(8) \triangleleft \Z_{3}, \ Spin(8) \triangleleft D_{6},\ SO(8),\ O(8),\ Oi(8),
\hfill\break PSO(8),\ PO(8),\ PSO(8) \triangleleft \Z_{3}, \ PSO(8) \triangleleft D_{6}$. 
\ss
\nt
Note that $Spin(4)= SU(2) \times SU(2),\ PSO(4)= SO(3) \times SO(3)$, and $Semispin(4) = SU(2) \times SO(3)$. 
\ss
These, then, are the non-trivial distinct groups corresponding to the $SO(n)$ algebra. When $n$ is 32, there are no fewer than eight candidates. String theory selects a particular group from among these eight in the following extraordinary way. In the heterotic theories, gauge fields arise in connection with the lattice of momenta on a sixteen-dimensional torus. The lattice must be even and self - dual. The crucial point is that these requirements impose conditions not merely on the root system of the gauge group, {\it but also on its integral lattice} [16]. However, there is a deep connection between the integral lattice and the global structure of a compact, connected Lie group. Thus string theory provides a route from strictly physical conditions directly to the global structure of the ( identity component of the ) gauge group. As is well known, $Semispin(32)$ satisfies these conditions, while $SO(32),\ Spin(32)$, and $PSO(32)$ do not. The argument is now completed by a glance at Theorem 1 : we see that $ Spin(32)$ and $PSO(32)$ each have a non-trivial disconnected version, and $SO(32)$ has two, but $Semispin(32)$ has none. The precise global structure of the gauge group is thereby fixed : it is $Semispin(32)$.
\ss
We close this section with some remarks on the representation theory of $Spin(4m),\hfill\break Semispin(4m), \ SO(4m)$, and $PSO(4m)$. Recall that the basic faithful representation of $ Spin(4m)$, obtained [14] by suitably restricting an irreducible representation of the Clifford algebra, has a canonical decomposition
$$ \triangle_{4m} = \triangle \ ^{+}_{4m} \oplus \triangle \ ^{-}_{4m},$$
\nt
where $\triangle \ ^{+}_{4m}$ is a representation with kernel $\{ 1,  \hat K_{4m} \}$ and $\triangle \ ^{-}_{4m}$ has kernel  $\{ 1, - \hat K_{4m} \}$. Thus, neither $\triangle \ ^{+}_{4m}$  nor $\triangle \ ^{-}_{4m}$ is faithful ; one must take their sum. Hence $\triangle \ ^{+}_{4m}$ and $\triangle \ ^{-}_{4m}$ are called the {\it half-spin representations}. Evidently they are faithful not on $Spin(4m)$ but on the group we (accordingly) call $Semispin(4m)$. The half-spin representations are of dimension $2^{2m-1}$. Thus the so-called `` 128-dimensional representation of $SO(16)"$ which plays a prominent role in string theory is in fact the defining representation of $Semispin(16)$. Again, the defining representation of $Semispin(32)$ is $32,768$ - dimensional, a decidedly inconvenient value. Fortunately, we have 
$$PSO(32) = Semispin(32) / \Z_{2},$$
\nt
and so every representation of $PSO(32)$ is automatically a representation of $Semispin(32)$; thus, the latter has a more manageable (but unfaithful) 496 - dimensional representation, which is also an unfaithful representation of  $SO(32)$ and $Spin(32)$, namely the adjoint. Similarly  
$PSO(16)$ yields a 120 - dimensional representation of $Semispin(16)$, and so the latter has a {\it faithful} 248 - dimensional representation defined by the direct sum, ${\bf 120 \oplus 128}$. As the representation is faithful, and as the (likewise faithful) adjoint of $E_{8}$ decomposes as ${\bf 248 = 120 \oplus 128}$, this immediately shows that $E_{8}$ contains $Semispin(16)$ and {\it not}, as is so often said, $SO(16)$. Thus ${(E_{8} \times E_{8}) \triangleleft \Z_{2}}$ has a maximal subgroup of the form  $(Semispin(16) \times Semispin(16))\triangleleft \Z_{2}$, and so we see that {\it the Semispin groups appear in both heterotic string theories}. In fact, Witten [17] has recently argued that the same is true of the Type I theory. The gauge group of Type I at the perturbative level is $PO(32)$ ( see Theorem 1). As this group is disconnected, while $Semispin(32)$ has no non-trivial disconnected version, this appears to obstruct the supposed $S$-duality between the Type I and the $``SO(32)"$ heterotic string theories [11]. However, Witten shows that a subtle non-perturbative effect breaks $PO(32)$ to $PSO(32) =  Semispin(32) / \Z_{2}$; furthermore, there appear to be Type I non-perturbative states transforming ``spinorially" under the gauge group. ( Note that, like \ a spinor, a ``semispinor" is odd under a $2 \pi$ rotation; the \ non-trivial \ element in \ the centre of $Semispin(32)$ is the projection of  $-1$ in  $Spin (32)$. ) In short, the gauge group of Type I string theory is undoubtedly  $Semispin(32)$ precisely, not $SO(32)$. The $Semispin$ groups appear in all three string theories with non-trivial gauge groups. 
\ss
All this appears to bode well for duality : in particular, since  ${(E_{8} \times E_{8}) \triangleleft \Z_{2}}$  contains $( Semispin(16) \times   Semispin(16)) \triangleleft \Z_{2}$, one would expect this same group to appear on the $Semispin
(32)$ side. In fact, this is {\it not} the case, as we now show. 
\bs

\nt
{\bf III. \ SUBGROUPS OF SEMISPIN(4m) CORRESPONDING TO }\hfill\break
${\bf SO(k) \times SO(4m-k)}$
\ss
Evidently $SO(32)$ contains $SO(16) \times SO(16)$ block - diagonally; more generally, $SO(4m)$ contains $SO(k) \times SO(4m-k), k \geq 2$. The product is indeed direct, since $SO(k)$ and 
$SO(4m-k)$ intersect trivially, in $ \{ I_{4m} \}$. However, $Spin(4m)$ {\it does not} contain  
$Spin(k) \times Spin (4m-k)$, because both factors contain $ \{ \pm 1 \}$. In fact, the subgroup is $Spin(k) \bullet Spin (4m -k)$, a {\it local direct product}, where

$$ Spin(k) \bullet Spin (4m-k) = (Spin(k) \times Spin (4m -k)) / Z_2,  $$ 
\nt
with $Z_2$ generated by $(-1, -1)$.
\ss
There is another important difference between $SO(4m)$ and $Spin (4m)$ in this area. It is clear that, when $k=2j$ is even,  $ Spin(k) \bullet Spin (4m -k)$ can be characterised as the group of all $Spin (4m)$ \  elements which \ commute with ( that is, as the {\it centraliser} \hfill\break of ) $\hat K_{2j}$. Now $\hat K_{2j}$ projects to the $SO(4m)$ matrix diag $( - I_{2j}, I_{4m - 2j})$ = $K_{2j}$, but the $SO(4m)$ centraliser of $K_{2j}$ is not $SO(2j) \times SO(4m - 2j)$; rather, it is the disconnected subgroup $S(O(2j) \times O (4m - 2j)),$ the set of all pairs $(A,B)$ in 
$O(2j) \times O (4m - 2j)$ such that $det \ A = det \ B$. That is, a Wilson loop that breaks $ Spin (4m)$  to a {\it connected} subgroup will break  $SO(4m)$  to a {\it disconnected} subgroup. ( Recall [2] that a Wilson loop in a gauge theory is a closed curve in space-time which has a non-trivial holonomy element even in the vacuum. The gauge group is broken to the centraliser of the (usually finite) subgroup generated by the holonomy element. ) 
\ss
Now we turn to the case of the $Semispin(4m)$. Suppose first that $k$ is odd. Then $ Spin(k) \bullet Spin (4m-k)$ does not contain  $ \hat K_{4m}$, and so it is unaffected by the projection from  $Spin (4m)$ to  $Semispin(4m)$. Thus, when $k$ is odd, the subgroup of $Semispin(4m)$ corresponding to $SO(k) \times SO(4m-k)$ is globally isomorphic to  $ Spin(k) \bullet Spin (4m-k)$. Next, suppose that $k=2j$ is even but not a multiple of $4$. Then $4m - 2j$ is likewise even but not a multiple of $4$, and so $Spin(2j)$ and $Spin(4m-2j)$ have $\Z_4$ centres generated respectively by  $\hat K_{2j}$  and $\hat K_{2j+1,4m}$. We have 
$$ \hat K_{2j} \hat K_{4m} = -\hat K_{2j+1,4m}$$
\nt
and 
$$\hat K_{2j+1,4m} \hat K_{4m} = -\hat K_{2j},$$
\nt
and so the effect of factoring by $ \hat K_{4m}$ is to identify the {\it entire} centre of $Spin (4m-2j)$ with that of $Spin (2j)$. We have, when $j$ is odd,   

$$Spin(2j)^{\bullet}_{\bullet}Spin(4m-2j) = (Spin(2j) \times Spin(4m-2j))/ \Z_4$$
\nt
as the subgroup of $Semispin(4m)$ corresponding to $SO(2j) \times SO(4m-2j)$. 
\ss
Finally, if $k=4j$ is a multiple of $4$, then so is $4m-4j$ and both $Spin(4j)$ and $Spin(4m-4j)$ have centres isomorphic to $\Z_{2} \times \Z_{2}$. These centres are $\{ \pm 1, \ \pm  \hat K_{4j} \}$ and $\{ \pm 1, \ \hat K_{4j+1,4m} \}$ respectively, and since we have 
$$ \pm \hat K_{4j} \hat K_{4m} = \pm \hat K_{4j+1,4m}$$ 
\nt
and 
$$\pm \hat K_{4j+1,4m} \ \hat K_{4m} = \pm \hat K_{4j},$$
\nt
we see that, once again, the effect of the projection  $Spin(4m) \longrightarrow Semispin(4m)$ is to identify the entire centre of  $Spin(4m-4j)$ with that of $Spin(4j)$. We use the notation 

$$ Spin(4j)^{\bullet}_{\bullet}Spin(4m-4j) = (Spin(4j) \times Spin(4m-4j))/(\Z_{2} \times \Z_{2}).$$
\ss
Next, recall that, from a physical point of view, we are interested in obtaining all these groups as centralisers of some element in $Semispin(4m)$. We saw earlier that the centraliser of 
$\hat K_{2j}$ in $Spin(4m)$ is connected, but that of $K_{2j}$ in  $SO(4m)$ is not. Let $K^{*}_{2j}$ be the projection of $ \hat K_{2j}$ to $Semispin(4m)$. The centraliser of $K^{*}_{2j}$ will include $Spin(2j)^{\bullet}_{\bullet} Spin(4m-2j)$, but it wil also include any $Semispin(4m)$ element $L^*$ such that $\hat L$, a lift of $L^*$ to  
$Spin(4m)$, satisfies $\hat L \ \hat K_{2j} = \hat K_{4m} \hat K_{2j} \hat L$. Projecting this to $SO(4m)$, we find that the corresponding matrices satisfy
$$L \ K_{2j} \ L^{-1} = -K_{2j},$$
\nt
whence Trace  $K_{2j} = 4(m-j)=0.$ Thus if $j \not= m, \ L^*$ does not exist, and so the centraliser of $K_{2j}^{\ast}$ in $Semispin(4m)$ is precisely $Spin(2j)^{\bullet}_{\bullet}  Spin(4m-2j)$. If 
$j=m$, then we have 
$$J_{2m} \ K_{2m} \ J_{2m}^{-1} = - K_{2m},$$
\nt
where
$$J_{2m} =  \left ( \matrix{0      & -I_{2m} \cr 
                            I_{2m} &  0      \cr}\right).$$
\nt             
This solution is essentially unique. The corresponding element of $Spin(4m)$ is (see Ref. [16], page 174, and modify suitably ) 

$$\hat J_{2m} = 2^{-m} (1 - e_{1} \ e_{1+{2m}}) (1 - e_{2} \ e_{2+2m}) \ldots  (1 - e_{2m} \ e_{4m}).$$ 
\ss
\nt
Now 
\nt
$$\hat K_{2m} \ \hat J_{2m}= 2^{-m} (e_{1} + e_{1+2m})(e_{2} + e_{2+2m}) \ldots (e_{2m} + e_{4m})$$  
\hskip4.55truecm  $= 2^{-m} (e_{1+2m} + e_{1})(e_{2+2m} + e_{2}) \ldots (e_{4m} + e_{2m})$  
\ss
\hskip3.68truecm  $= \hat J_{2m} \hat K_{1+2m,4m} = (-1)^{m} \hat K_{4m} \hat J_{2m} \hat K_{2m}$
\ss
\nt
since  $(\hat K_{2m})^{2} = (-1)^{m}$ and  $\hat K_{2m} \hat K_{1+2m,4m} = \hat K_{4m}$. Thus if $m$ is even, the projections to $Semispin (4m)$ satisfy   $K_{2m}^{\ast} J_{2m}^{\ast} = J_{2m}^{\ast} K_{2m}^{\ast}$ as required. If $m$ is odd, we project instead to  $Spin(4m) / \{1, - \hat K_{4m} \}$ and recall that this is isomorphic to $Semispin (4m)$. A further exercise in Clifford algebra shows that \hfill\break
$$ (\hat J_{2m})^{2} = (-1)^{m}  \hat K_{4m},$$
\nt
and so the appropriate projections are of order two. The effect on $Spin(2m)^{\bullet}_{\bullet}  Spin(2m)$ of conjugation by $J_{2m}^{\ast}$ is to exchange the two factors. We conclude that {\it the centraliser of} $K_{2j}^{\ast}$ {\it in} $Semispin(4m)$ {\it is} 

$$Spin(2j)^{\bullet}_{\bullet} Spin(4m-2j)\  {\rm if}\ j \not= m$$
$$(Spin(2m)^{\bullet}_{\bullet} Spin(2m)) \triangleleft \Z_{2} \ {\rm if} \ j= m.$$
\ss
Finally, let us consider the specific case of $Semispin(32)$. Its $``SO(16) \times SO(16)"$ subgroup is actually  $(Spin(16)^{\bullet}_{\bullet} Spin(16)) \triangleleft \Z_{2}$ where the full centre of each $Spin(16)$ is identified with that of the other, and where $\Z_{2}$ exchanges the two factors. Compare this with the $``SO(16) \times SO(16)"$ subgroup of $(E_{8} \times E_{8}) \triangleleft \Z_{2}$, which is $(Semispin(16) \times Semispin(16)) \triangleleft \Z_{2}$. Both groups have the ``exchange" $\Z_{2}$, which is welcome from the point of view of $T$-duality. But $Spin(16)^{\bullet}_{\bullet} Spin(16)$ {\it is not isomorphic to} $Semispin(16) \times Semispin(16)$. Both are $\Z_{2} \times \Z_{2}$ quotients of $Spin(16) \times Spin(16)$, but 
$\Z_{2} \times \Z_{2}$  acts differently in each case. This implies that neither is a cover of the other, and so they each have representations which cannot be regarded as representations of the other. For example, by factoring out $\{ \pm 1 \}$ in $ Spin(16)^{\bullet}_{\bullet} Spin(16)$, we obtain $SO(16) \bullet SO(16)$, where the dot means that the two factors have a non-trivial intersection, $\{ \pm \ I_{16} \}$. The tensor product of the vector with itself, $( {\bf 16, 16 })$, is faithful for this group, and so it is a representation of 
$Spin (16) ^{\bullet}_{\bullet} Spin (16)$. This representation contains faithful copies of $SO(16)$, something which is impossible for any representation of $Semispin (16) \times  Semispin (16)$. On the other hand, let $({\bf 128, 1}) \oplus  ({\bf 1, 128})$ be the defining representation of $Semispin (16) \times  Semispin (16)$; this representation distinguishes the centres of the two  $ Semispin (16)$ factors, which cannot happen in any representation of $Spin (16) ^{\bullet}_{\bullet} Spin (16)$. ( If we take the quotient of $ Spin (16)^{\bullet}_{\bullet}  Spin (16)$ by $ \{ 1, \hat K_{16} \}$, then we obtain  $ Semispin (16) \bullet   Semispin (16)$, in which the two factors intersect in  $\{ \pm 1 \}$, where we use $-1$ to denote the projection of $-1$ in $ Spin(16)$; but $( {\bf 128, 1}) \oplus  ( {\bf 1, 128})$ does not descend to a representation of this group. ) The two groups do have some representations in common, such as $( {\bf 128, 128})$ and $( {\bf 120, 1}) \oplus  ( {\bf 1, 120})$, the latter being the defining representation for $PSO (16) \times PSO (16)$, which is a $\Z_{2} \times \Z_{2}$ quotient of both $Semispin (16) \times  Semispin (16)$ and $Spin (16)^{\bullet}_{\bullet} Spin (16)$. However, while $( {\bf 120, 1}) \oplus  ( {\bf 1, 120})$ is important for duality, so also are  $( {\bf 16, 16})$ and  $( {\bf 128, 1}) \oplus  ( {\bf 1, 128})$. 
\ss
We see, then, that the appearance of $Semispin$ groups in both heterotic theories was somewhat deceptive; for $Semispin(4m)$ is strangely unlike $Spin(4m)$ and $SO(4m)$. While these last contain subgroups of the same kind as themselves, $Spin(2j)^{\bullet}_{\bullet} Spin (4m-2j)$ and $SO(2j) \times SO(4m-2j)$ respectively, $Semispin(4m)$ ${\it does \ not \ contain} $ $Semispin(2j) \hfill\break \bullet Semispin (4m-2j)$. Instead, it contains  $ Spin(2j)^{\bullet}_{\bullet} Spin (4m-2j)$. Thus we arrive at the disconcerting fact that while $E_{8} \times E_{8}$ contains  $Semispin$ groups,  $ Semispin(32)$ itself does not. The ``common  $SO(16) \times SO(16)$ subgroup " which appears in the duality literature not only fails to be isomorphic to  $SO(16) \times SO(16) {\bf : }$ it simply {\it does not exist}. 
\ss
One way to approach this problem is to find a group which covers both $Semispin(16) \times  Semispin(16)$ and $Spin(16)^{\bullet}_{\bullet} Spin(16)$, since all of the representations of {\it both} groups will then be representations of that group. One obvious choice is $Spin(16) \times Spin(16)$, but there is a better alternative, constructed as follows. Write $Spin(16) \times Spin(16)$ as $Spin(16)^{L} \times Spin(16)^{R}$ and let $ \{ \pm 1^{L}, \pm \hat K_{16}^{L} \}$,  $ \{ \pm 1^{R}, \pm \hat K_{16}^{R} \}$ be the respective centres. We define   
$$Spin(16) \ast Spin(16)= (Spin(16) \times Spin(16)) / \{( 1^{L}, 1^{R}), (\hat K_{16}^{L}, \hat K_{16}^{R}) \}.$$
\ss
That is, we identify $\hat K_{16}^{R}$ with $\hat K_{16}^{L}$. Further factoring by this element produces $Semispin(16) \times  Semispin(16)$, while factoring by $(-1^{L}, -1^{R})$ produces $Spin(16)^{\bullet}_{\bullet} Spin(16)$. That is,  $Spin(16) \ast Spin(16)$ is a double cover of ${\it both}$  $Semispin(16) \times  Semispin(16)$ {\it and} $Spin(16)^{\bullet}_{\bullet} Spin(16)$. Hence $(\bf {16,16})$, $( {\bf 128, 1}) \oplus  ( {\bf 1, 128})$, and $( {\bf 120, 1}) \oplus  ( {\bf 1, 120})$ are all (unfaithful) representations of  $Spin(16) \ast Spin(16)$, and   $(\bf {16,16}) \oplus ({\bf 128, 1}) \oplus ( {\bf 1, 128})$ is a ${\it faithful}$ representation of   
$Spin(16) \ast Spin(16)$, though it is not a representation of  $Semispin(16) \times  Semispin(16)$ or  $Spin(16)^{\bullet}_{\bullet} Spin(16)$, much less $SO(16) \times  SO(16)$. 
(It is faithful because $\hat K_{16}$, the one non-trivial element of $Spin(16) \ast Spin(16)$ mapped to the identity by $({\bf 128, 1}) \oplus  ({\bf 1, 128})$, acts as $-1$ in $( {\bf 16, 16}).)$
\ss
In fact, $Spin(16) \ast Spin(16)$ is the gauge group of the unique tachyon-free 10 dimensional non-supersymmetric heterotic string theory [18],[19], which plays a central role in recent investigations of strong-coupling duality [20]. The massless spectrum of this theory consists of a gravity multiplet, spacetime vectors assigned to $({\bf 120, 1}) \oplus ({\bf 1, 120})$, and spacetime spinors assigned to the ``$SO(16) \times  SO(16)$" representation $({\bf 16, 16}) 
\oplus ({\bf 128, 1}) \oplus ( {\bf 1, 128})$, which, as we have seen, is a faithful representation of $Spin(16) \ast Spin(16)$.
\ss
We claim, then, that the string theorist's $``SO(16) \times  SO(16)"$ is actually $Spin(16) \ast Spin(16)$. The strange feature of this conclusion is that $Spin(16) \ast Spin(16)$ {\it is not a subgroup of either} $E_{8} \times E_{8}$ {\it or} $Semispin(32)$. (Nor can it be embedded in $Spin(32), SO(32)$, or $PSO(32))$. Thus {\it it does not make sense} to speak of breaking   
$E_{8} \times E_{8}$ or $Semispin(32)$ to $Spin(16) \ast Spin(16)$ by a Wilson loop or in any other way. We believe that the way to solve this problem is through a study of ``generalised Stiefel-Whitney classes" [9],[10]. For example, to establish the duality of a certain $Semispin (32)$ configuration with an $(E_{8} \times E_{8}) \triangleleft \Z_{2}$ configuration, one breaks $Semispin(32)$ to $(Spin(16)^{\bullet}_{\bullet} Spin(16)) \triangleleft Z_{2}$, lifts this to a $(Spin(16) \ast Spin(16))\triangleleft \ \Z_{2}$ structure ( checking that the appropriate generalised Stiefel-Whitney class vanishes ), projects this to a $(Semispin (16) \times  Semispin (16)) \triangleleft \Z_{2}$ structure, and then extends to $(E_{8} \times E_{8}) \triangleleft  \Z_{2}$ The details of this process will be described elsewhere. 
\ss
Let us summarise as follows.\ $SO(4m), m > 2$, has important subgroups of the form $SO(k) \times SO(4m-k)$, though in fact this is just the identity component of $S(O(k) \times O(4m-k))$. The other three connected groups locally isomorphic to $SO(4m)$, namely 
$Spin(4m), Semispin (4m)$,and $PSO(4m)$, have analogous subgroups given by the following Theorem. 
\ss
\nt
THEOREM 2: The Lie algebra inclusion ${\cal SO}(k) \oplus {\cal SO}(4m-k) \rightarrow  {\cal SO}(4m)$ has the following counterparts at the Lie group level. 
\ss
\nt
$S(O (k) \times O(4m-k)) \rightarrow SO(4m)$ \hfill\break
$Spin(k) \bullet Spin(4m-k)  \rightarrow Spin(4m)$ \hfill\break
$PS(O (k) \times O(4m-k)) \rightarrow PSO(4m)$ \hfill\break
$Spin(k) \bullet Spin(4m-k)  \rightarrow Semispin(4m) \; (k \ odd) $ \hfill\break
$Spin(2j) ^{\bullet}_{\bullet} Spin(4m-2j) \rightarrow Semispin (4m) \; (k=2j,j \not = m) $ \hfill\break
$(Spin(2m)^{\bullet}_{\bullet} Spin(2m)) \triangleleft \Z_{2} \rightarrow Semispin (4m)$ \hfill\break
\ss
\nt
Here a single dot denotes a factoring by a diagonal $\Z_{2}$, as also does the prefix $P$, while the double dot denotes a factoring by a diagonal $\Z_{4}$ or by $\Z_{2} \times \Z_{2}$
as the case may be. 
\ss
 
\vfill\eject
\bs
\nt
{\bf IV SUBGROUPS OF SEMISPIN (4m) CORRESPONDING TO U(2m)}.
\ss
Another subgroup of $SO(4m)$ which plays an important role in the string literature ( see, for example, Refs [9],[20],[21] ) is the unitary group $U(2m)$. If $A + {\it i}B$ is any element of $U(2m)$, where $A$ and $B$ are real, then 
$\left( \matrix{A & -B \cr 
                 B & {\phantom{-}} A \cr}\right)$ is an element of $SO(4m)$. The unitary subgroup can also be characterised as the centraliser of the matrix $J_{2m}$ defined in the preceding section. ( Notice that $J_{2m}$ is $SO(4m)$-conjugate to diag 
$\left (\pmatrix {0 & -1   \cr 
                 1 & \phantom {-} 0 \cr}, 
        \pmatrix {0 & -1   \cr 
                  1 & \phantom {-} 0  \cr}\ldots \right )$, and the reader can take $J_{2m}$ to be defined in this way if that is convenient. ) 
\ss
Now $U(2m)$ is {\it not} isomorphic to $U(1) \times SU(2m)$, because $U(1)$ and  $SU(2m)$ intersect non-trivially. Let $z$ be a primitive $(2m)$-th root of unity, and let $\Z_{2m}$ act on $U(1) \times SU(2m)$ by
$$(u,s) \rightarrow (uz^{-1},zs).$$
Then $(U(1) \times SU(2m))/ Z_{2m}$ is isomorphic to $U(2m)$. Elements of $U(2m)$ may therefore be represented as equivalence classes, $[u,s]_{2m}$. 
\ss
Now we ask : what is the subgroup of $Spin(4m)$ which projects onto $U(2m)$? It is useful to notice that the answer cannot be isomorphic to $U(2m)$, for $U(2m)$ would be of maximal rank in 
$Spin(4m)$, and so the centre of $Spin(4m)$ would be contained in the centre of $U(2m)$; that is, we would have  $\Z_{2} \times \Z_{2}$ contained in $U(1)$, which is impossible. ( This argument would not work for the $U(2m+1)$ subgroup of $SO(4m + 2)$, and indeed the cover of  $U(2m+1)$ in $Spin(4m+2)$ is again isomorphic to $U(2m+1).)$ In fact, it is not difficult to see that $Spin(4m)$ contains $(U(1) \times SU(2m))/ \Z_{m}$, which consists of pairs $[u,s]_{m}$. In this group, $[z^{-1}, z \ I_{2m}]_{m}$ is not the identity, though $[z^{-1}, z \ I_{2m}]_{2m}=1$; therefore $[z^{-1}, z \ I_{2m}]_{m}$ corresponds to $-1$ in $Spin(4m)$. There is another important element of order two in this group,  $[1, -I_{2m}]_{m}$, but of course there are others, such as $[z^{-1}, -z \ I_{2m}]_{m}$. In order to determine the structure of the subgroup of $Semispin(4m)$ corresponding to $U(2m)$, we must determine which of these corresponds to $\hat K_{4m}$.  
\ss
\nt
THEOREM 3 : Let $SemiU(2m)$ denote the projection of $(U(1) \times SU(2m) / \Z_{m}$ to $Semispin(4m)$. Then the global structure of $SemiU(2m)$ is given as follows : \hfill\break
\nt
\ss
$SemiU(2m) = [ U(1) \times (SU(2m)/ \Z_{2})]/ \Z_{m/2}, \ m \ {\rm even}$

\hskip2.3truecm = $[U(1) \times SU(2m)] / \Z_{m},\  m$ odd.
\ss
\nt
PROOF: Under the embedding of $U(2m)$ in $SO(4m)$, the matrix $J_{2m}$ arises from the 
$U(2m)$  matrix ${\it i}I_{2m}$, which is $[{\it i},I_{2m}]_{2m}$. Thus we see that the $Spin(4m)$ element  $\hat J_{2m}$ defined in the preceding section must be either $[{\it i},I_{2m}]_{m}$ or $[{\it i}z^{-1}, z \ I_{2m}]_{m}.$ In either case we have $(\hat J_{2m})^{2}= [-1, I_{2m}]_{m}$. Now recall that $(\hat J_{2m})^{2} = (-1)^{m} \ \hat K_{4m}$ and that we have agreed to define $Semispin(4m)$ by $Spin(4m) / \{1,(-1)^{m} \hat K_{4m} \}$ for convenience, so that $J_{2m}^{\ast}$ is always of order two. ( See the remarks at the end of this section. ) Thus when $m$ is odd, we must factor by
$$- \hat K_{4m} = [-1, I_{2m}]_{m} \not= [1, -I_{2m}]_{m}.$$
Clearly, the factoring will affect $U(1)$ but not $SU(2m)$. However, $U(1) / \Z_{2}= U(1)$, since the map $u \rightarrow u^{2}$ is a group epimorphism for this infinite abelian group. Thus we obtain $[U(1) \times SU(2m)]/ \Z_{m}$ when $m$ is odd. When $m$ is even, $\hat K_{4m}$ is 
$[-1, I_{2m}]_{m}$, which is equal to $[1, -I_{2m}]_{m}$. The factoring will affect both $U(1)$  and $SU(2m)$ in this case, and, after it, the $\Z_{2}$ in $\Z_{m}$ will act trivially; so we obtain $(U(1)/ \Z_{2}) \times (SU(2m)/ \Z_{2})$, with an effective action by $\Z_{m/2}$. Hence the group is  $[(U(1) \times (SU(2m)/ \Z_{2})] / \Z_{m/2}$, and this completes the proof. 
\ss
Notice that, according to this theorem,
$$SemiU(2) = [U(1) \times SU(2)]/ \Z_{1}= SO(2) \times SU(2),$$
\nt
which is indeed a subgroup of $Semispin(4) = SO(3) \times SU(2)$. Again, 
$$SemiU(4) = [U(1) \times  (SU(4)/ \Z_{2})] / \Z_{1}= SO(2) \times SO(6),$$
\nt
which is contained in $SO(8)= Semispin(8).$ 
\ss
Clearly  $SemiU(2m)$  is the identity component of the centraliser, in  $Semispin(4m)$, of $J_{2m}^{\ast}$, the projection of $\hat J_{2m}$. Recall that, unlike  $J_{2m}$ in $SO(4m)$ and  $\hat J_{2m}$ in $Spin(4m)$ (which are both of order 4), $\hat J_{2m}^{\ast}$ is of order 2; this is important for applications [9],[21]. For example, consider a $Semispin(32)$ heterotic theory compactified on a $K3$ surface which is a Kummer surface at an orbifold limit, with a point - like instanton at the singular point [21]. Excising this point, we obtain a neighbourhood which retracts to the projective sphere, $S^{3}/Z_{2}$. If $J_{16}^{\ast}$  were of order four, then it could not be realised as a holonomy element over $S^{3} / \Z_{2}$, and so the gauge group would not break. But $J_{16}^{\ast}$ is of order two, and so it {\it can } be realised as a holonomy over $S^{3}/Z_{2}$. (In the literature it is always {\it assumed } that a finite group $F$ can always be realised as a holonomy group over manifolds of the form $M/F$. This is true - with very mild conditions - but not at all obvious [6].) Then (unless one arranges to avoid it [21]) $Semispin(32)$ will break to the centraliser of 
$J_{16}^{\ast}$  in $Semispin(32)$. This includes $SemiU(16)$, but it also includes $K_{16}^{\ast}$, as we saw in the preceding section. The matrix $K_{16}= {\rm diag} (-I_{16},I_{16})$ acts by complex conjugation on U(16), that is, 
$K_{16} \left ( \matrix {A & -B   \cr 
                          B & \phantom {-} A \cr}\right)$ 
$K_{16}^{-1} = \left ( \matrix {\phantom {-}A & B   \cr 
                                           -B & A \cr}\right)$. Similarly, conjugation by  $K_{16}^{\ast}$ maps elements of $SemiU(16)$ to their complex conjugates. ( A typical element of $SemiU(16) = [U(1) \times (SU(16) / \Z_{2})] / \Z_{4}$ has the form $[u^{2},[s]_{2}]_{4}$, with complex conjugate $[(\bar u^{2}, [\bar s]_{2}]_{4}$; bear in mind that $\Z_{4}$ acts on  
 $U(1) \times (SU(16) / \Z_{2})$ by $(u^{2},[s]_{2}) \rightarrow ({\it i}u^{2},[zs]_{2})$, with 
$z^{8}=1.)$ As $(K_{16}^{\ast})^{2}=1$, we see that $J_{16}^{\ast}$ breaks $Semispin(32)$ to $SemiU(16) \triangleleft \Z_{2}$ and ${\it not }$ [9],[21] to U(16) or $U(16)/ \Z_{2}$, which are quite different to $SemiU(16)$. Notice the differences with $SO(32): J_{16}$ is of order four, and its centraliser in $SO(32)$ is the connected group $U(16)$, while $J_{16}^{\ast}$ is of order two, with $Semispin(32)$ centraliser isomorphic to the {\it disconnected} group $SemiU(16) \triangleleft \Z_{2}$. On the other hand, $J_{16}$ cannot break $SO(32)$ over $S^{3}/ \Z_{2}$, but $J_{16}^{\ast}$  can break $Semispin(32)$.\hfill\break ( Note also that there do exist bundles over $S^{3}/ \Z_{2}$ having the full disconnected group $SemiU(16) \triangleleft \Z_{2}$ as holonomy group [6].)
\ss
In the same way, $J_{8}^{\ast}$ is of order two, and it breaks $Semispin(16)$ to $SemiU(8) \triangleleft \Z_{2}$, with $Z_{2}$ generated by $K_{8}^{\ast}$, and with $SemiU(8) = [U(1) \times (SU(8) / \Z_{2})]/ \Z_{2}$. However, $Semispin(16)$ is mainly of interest because it is a maximal subgroup of $E_{8}$. If $J_{8}^{\ast}$ is embedded in $E_{8}$ through $Semispin(16)$, then of course its centraliser must contain $SemiU(8)\triangleleft \Z_{2}$; however, this cannot be the full centraliser, since the centraliser of any element of a simply connected compact Lie group (such as $E_{8}$, but not $Semispin(16)$) must be connected. Hence the centraliser of  $J_{8}^{\ast}$ must be a connected subgroup of $E_{8}$ containing $SemiU(8) \triangleleft \Z_{2}$. Of course, $Semispin(16)$ is such a subgroup, but there is another. The exceptional Lie group $E_{7}$ has a maximal rank subgroup [22] isomorphic to $SU(8)/ \Z_{2}$, and in fact one can prove that $E_{7}$ contains a disconnected subgroup with two connected components, one being $SU(8)/ \Z_{2}$. Combining this with a Pin(2) subgroup of $SU(2)$, we obtain, after suitable identifications, $SemiU(8)\triangleleft \Z_{2}$ as a subgroup of $SU(2) \bullet E_{7}$, which is a maximal subgroup of  $E_{8}$. In fact, the centraliser of  $J_{8}^{\ast}$  in $E_{8}$ is $SU(2) \bullet E_{7}$, while that of $-J_{8}^{\ast}$  turns out to be just $Semispin(16)$; this is important in applications [9].
\ss
One of the most interesting and important applications where the distinction between $SO(4m)$ and $Semispin(4m)$ is crucial concerns $K3$ compactifications of the $(E_{8} \times  E_{8})\triangleleft \Z_{2}$ heterotic theory. When the instanton numbers are assigned symmetrically to the two factors, one finds [9] that the corresponding (T-dual) $``SO(32)"$ configuration corresponds to a $Semispin(32)$ bundle which {\it does not} lift to a $Spin(32)$ bundle. This is the Semispin analogue of the failure of the orthonormal frame bundles over certain Riemannian manifolds [14] to lift to spin bundles. If a $Semispin(32)$ bundle does lift to a $Spin(32)$ bundle, then it will automatically define ( by projection ) an $SO(32)$ bundle; and so a $Semispin(32)$ bundle which fails to lift to a $Spin(32)$ bundle is said to lack a ``vector structure".
\ss
Examples of such $Semispin(4m)$ bundles can be given by once again exploiting the fact that $J_{2m}^{\ast}$ is of order two, whereas $\hat J_{2m}$, its counterpart in $Spin(4m)$, satisfies 
$(\hat J_{2m})^{2} = (-1)^{m} \hat K_{4m}$ and so is of order four (like $J_{2m}$). This makes it possible to construct a non-trivial U(1) bundle over a two - cycle in the base, such that connections on this bundle satisfy the usual (``Dirac") integrality conditions, but their pull-backs to a covering bundle would not. When this U(1) bundle is extended to a $Semispin(32)$ bundle, therefore, the latter cannot be lifted to a double cover. It is in precisely this way that the dual partner of the above [9] $(E_{8} \times E_{8})\triangleleft \Z_{2}$  compactification is constructed. One could not wish for a more striking confirmation of the importance of the distinction between $SO(4m)$ and $Semispin(4m)$. 
\ss
In this spirit, \ we ask whether \ $U(1)$ is indeed the precise \ global form of the gauge group in question. This $U(1)$ may be identified as the explicit U(1) in $[U(1) \times \hfill\break
(SU(16) / \Z_{2})]/ \Z_{4}$, but we know that this group is most naturally regard as the identity component of $SemiU(16) \triangleleft \Z_2$. Therefore one should really regard the canonical $U(1)$ in $Semispin(32)$ as the identity component of $U(1) \triangleleft \Z_2$ or $O(2)$ in the notation of Theorem 1. ( The corresponding subgroup of SO(32) consists of all $32 \times 32$ matrices of the form 
$ \left (\matrix {I_{16} \cos \theta & -I_{16} \sin  \theta  \cr 
                   I_{16} \sin  \theta & I_{16} \cos \theta \cr}\right)$ 
$ \left( \matrix {-I_{16} \cos \theta & I_{16} \sin  \theta  \cr 
                    I_{16} \sin \theta & I_{16} \cos \theta \cr}\right)$. )
More generally, Semispin(4m) has a canonical subgroup of the form 

$$U(1) \bigcup K_{2m}^{\ast} \bullet U(1),$$
\nt
where U(1) corresponds to the Lie algebra element $\left(\matrix {0 & -I_{2m}   \cr 
                                                                  I_{2m} & 0 \cr}\right)$, or to its conjugate diag 
$\left (\pmatrix {0 & -1   \cr 
                 1 & \phantom {-} 0 \cr}, 
        \pmatrix {0 & -1   \cr 
                  1 & \phantom {-} 0  \cr}, \ldots \right )$
if one prefers. Recalling that $(K_{2m}^{\ast})^{2} = (-1)^{m}$, we see that the global structure is $O(2)$ if $m$ is even, but $Pin(2)$ if $m$ is odd ( see Theorem 1 ). One can actually prove that Semispin(4m) has no 
$Pin(2)$ subgroup containing U(1) when $m$ is even, and no $O(2)$ subgroup containing $U(1)$ when $m$ is odd. 
\ss
In constructing $Semispin(4m)$ bundles without ``vector structure", then, one should really begin with non-trivial $O(2)$ or $Pin(2)$ bundles. ( Of course, if the base manifold is simply connected, such a bundle will always reduce to a $U(1)$ bundle, but realistic string compactifications are not likely to be simply connected. ) Now we know that a $U(1)$ instanton breaks $Semispin(4m)$ to $SemiU(2m)$; what is the corresponding subgroup for $O(2)$ or $Pin(2)$ ? Since $K_{2m}^{\ast}$ acts by complex conjugation on all of  $SemiU(2m)$, the answer is the real subgroup of $SemiU(2m)$. The real subgroup of $SU(2m)$ is $SO(2m)$, while that of $SU(2m)/ \Z_{2}$ is $PSO(2m)$, and $U(1)$ contributes $J_{2m}^{\ast}$; finally, $-1$, the central element of $Semispin(4m)$, must of course also be included. Thus an $O(2)$ or $Pin(2)$ instanton will break $Semispin(4m)$ to 
$$\Z_{2} \times \Z_{2}  \times PSO(2m), \hskip2.0truecm {\rm m \; even},$$
$$\Z_{2} \times \Z_{2}  \times SO(2m), \hskip2.2truecm {\rm m \; odd},$$
\nt
with one $\Z_{2}$ generated by $-1$ and the other by $J_{2m}^{\ast}$. In particular,then, an $O(2)$ instanton in a $Semispin(32)$ theory will reveal itself by the presence of $PSO(16)$ where $SemiU(16)$ might be expected. (Note that this same PSO(16) arises in the $Spin(16) \hfill\break ^{\bullet}_{\bullet} Spin(16)$ subgroup of $Semispin(32)$, as the diagonal subgroup.)
\ss
Before concluding this section, we draw the reader's attention to the following point. While it is true that ${\rm Spin}(4m)/ \{1, - \hat K_{4m} \}$ is isomorphic to ${\rm Spin}(4m)/ \{1,  \hat K_{4m} \}$, the isomorphism is through an outer automorphism of $Spin (4m)$ which can change the way in which a given sub-algebra is embedded in the algebra of $Spin(4m)$, and this in turn can affect the global structure of the subgroup to which that sub-algebra exponentiates. A simple example is provided by $Spin(4) = SU(2) \times SU(2)$. Obviously  $SU(2) \times SO(3)$
is isomorphic to $SO(3) \times SU(2)$, but it is true that a given, fixed $SU(2)$ algebra exponentiates either to $SU(2)$ or to $SO(3)$, depending on whether one factors by $\{1,  \hat K_{4} \}$ or $ \{1, - \hat K_{4} \}$. We have chosen to define $Semispin(4m)$ by factoring $ \{1, - \hat K_{4m} \}$ when $m$ is odd, but one could decide to factor by $ \{1, \hat K_{4m} \}$, though in that case ${J_{2m}^{\ast}}$ will not commute with ${K_{2m}^{\ast}}$ and it will not be of order two. If one does this, then $[U(1) \times SU(2m)]/ \Z_{m}$ no longer projects to $[(U(1)/\Z_{2}) \times SU(2m)]/ \Z_{m}$. Instead we have 
$$ \hat K_{4m} = -(\hat J_{2m})^{2} = [-z^{-1}, z \ I_{2m}]_{m},$$
\nt
where $z$ is a primitive $(2m)$-th root of unity. This gives us $\hat K_{4m} = [z^{m-1}, z \ I_{2m}]_{m} = [1, z^{m} \ I_{2m}]_{m}$, because, since $m$ is odd, $m-1$ is even. Thus in fact 
$\hat K_{4m} = [1, -I_{2m}]_{m}$, and so the quotient of $(U(1) \times SU(2m))/ \Z_{m}$ by $\{1, 
\hat K_{4m} \}$ is isomorphic to $[U(1) \times (SU(2m)/ \Z_{2})]/ \Z_{m}$, which is not isomorphic to  $[U(1)/ \Z_{2} \times SU(2m)]/ \Z_{m}.$ Thus there is no unique subgroup of $Semispin(4m)$ corresponding to $U(2m)$ unless one specifies precisely which projection from $Spin(4m)$ to $Semispin(4m)$ one proposes to use. 
\ss
Our point of view is that for physical applications it is important that  $J_{2m}^{\ast}$ should be of order two rather than, like $J_{2m}$ and $\hat J_{2m}$, of order four. This is stressed repeatedly, for example, in Ref.[9]. This fixes the projections : we must factor out $\{1, 
\hat K_{4m} \}$ when $m$ is even, and $\{1, -\hat K_{4m} \}$ when $m$ is odd. 
\bs
\nt
{\bf  V. SUBGROUPS OF SEMISPIN(4m) CORRESPONDING TO} ${\bf Sp(1) \bullet Sp(m)}$.
\ss
Another subgroup of $SO(4m)$ which is important in various applications (see, for example, Ref. [21]) is the symplectic group $Sp(m)$, which embeds through $Sp(1) \bullet Sp(m)$. The latter has the global structure $[Sp(1) \times Sp(m)]/ \Z_{2}$. The corresponding subgroup of $Semispin(4m)$ is given as follows. 
\ss
\nt
THEOREM 4 : The Lie algebra inclusion ${\cal S}{\mit p}(1) \oplus {\cal S}{\mit p}(m) \rightarrow {\cal SO}(4m)$ has the following counterparts at the Lie group level. 
\ss
\nt
$Sp(1) \bullet Sp(m) \rightarrow SO(4m)$\hfill\break
\nt
$Sp(1) \times Sp(m)  \rightarrow Spin(4m)$  \hskip2.8truecm $m$ odd \hfill\break
\nt
$Sp(1) \bullet Sp(m) \rightarrow Spin(4m)$  \hskip3.0truecm $m$ even \hfill\break
\nt
$SO(3) \times PSp(m) \rightarrow PSO(4m)$ \hfill\break
\nt
$SO(3) \times Sp(m) \rightarrow Semispin(4m)$ \hskip1.9truecm $m$ odd \hfill\break
\nt
$SO(3) \times PSp(m) \rightarrow Semispin(4m)$ \hskip1.6truecm $m$ even 
\ss
\nt
Here a dot denotes a factoring by a diagonal $\Z_{2}$, and $PSp(m) = Sp(m)/ \Z_{2}$. 
\ss
\nt
PROOF: Consider first the case of $Spin(4m)$. We know that $Spin(4m)$ contains a subgroup of the form $[U(1) \times SU(2m)]/ \Z_{m}$, consisting of pairs $[u,s]_{m}$. Evidently we have 
$$[-1, I_{2m}]_{m} = [1, -I_{2m}]_{m}$$
\nt
when $m$ is even, but not when $m$ is odd. That is, the $\Z_{2}$ in $U(1)$ is identified with the $\Z_{2}$ in $SU(2m)$ when $m$ is even, but not when $m$ is odd. However, this $U(1)$ is contained in $Sp(1)$, and $SU(2m)$ contains $Sp(m)$; furthermore the central $\Z_{2}$ in $Sp(1)$ is the $\Z_{2}$ in $U(1)$, and the central $\Z_{2}$ in $Sp(m)$ is identical to the $\Z_{2}$  in $SU(2m)$. Thus we see that the central $\Z_{2}$ in $Sp(1)$ is identified, in $Spin(4m)$, with the central  $\Z_{2}$ in $Sp(m)$, if and only if $m$ is even. Hence the group is $Sp(1) \times Sp(m)$ if $m$ is odd, but $Sp(1) \bullet Sp(m)$ if $m$ is even. 
\ss
We saw, in the proof of Theorem 3, that $(-1)^{m} \hat K_{4m}= [-1, I_{2m}]_{m}$, the generator of the $\Z_{2}$ in $Sp(1)$. Hence $Sp(1) \times Sp(m)$ projects to $(Sp(1)/ \Z_{2}) \times Sp(m)$ when $m$ is odd, while $Sp(1) \bullet Sp(m)$ projects to $(Sp(1)/ \Z_{2}) \times (Sp(m)/\Z_{2})$ when $m$ is even. Recalling  that $Sp(1)/ \Z_{2} = SO(3)$ and $Sp(m)/ \Z_{2} = PSp(m)$, we have the stated results. Similarly, in $SO(4m)$, the central $Z_{2}$ coincides with the $Z_{2}$ in $Sp(1)$  and $Sp(m)$, so taking the quotient throughout $Sp(1) \bullet Sp(m) \rightarrow SO(4m)$, we obtain $SO(3) \times PSp(m) \rightarrow PSO(4m)$. This completes the proof. 
\ss
Notice that the theorem asserts that $SO(3) \times Sp(1)$ is contained in $Semispin(4)$, which is correct since the latter is $SO(3) \times SU(2)$ and $SU(2) = Sp(1)$. It also asserts that $Semispin(8) = SO(8)$ contains $SO(3) \times PSp(2)$, which is correct since $Sp(2) = Spin(5)$ and so $PSp(2)= SO(5)$. The theorem gives us $SO(3) \times PSp(8)$ as the subgroup of $Semispin(32)$  corresponding to $Sp(1) \bullet Sp(8)$ in $SO(32)$; this agrees with Ref. [21], where the importance of the  $SO(3)$ factor, appearing unexpectedly as a subgroup of $Semispin(32)$, is explained. ( Note that the full cover of $Sp(1) \bullet Sp(m)$ in $Spin(4m), m$ even, is actually $\Z_{2} \times Sp(1) \bullet Sp(m)$, where $\Z_{2} = \{ \pm 1 \}$; this projects to $\Z_{2} \times SO(3) \times PSp(m)$, so one might give this as the correct subgroup of $Semispin(4m).)$
\ss
\nt
{\bf VI. CONCLUSION}
\ss
The Semispin groups are of fundamental importance in string theory. The gauge groups of Type I and one of the heterotic theories are both $Semispin(32)$ precisely, while $E_8$ contains $Semispin(16)$; the latter in turn contains (see Theorem 2) $Spin(6) ^\bullet_\bullet Spin(10) \hfill\break= [SU(4) \times Spin(10)]/ \Z_{4}$, and so it provides a possible route to $Spin(10)$ grand unification. 
\ss
These facts alone warrant a detailed study of the Semispin groups and their remarkable subgroups. Theorem 1 places $Semispin(4m)$ in the context of the entire family of non-trivial groups locally isomorphic to $SO(n)$, while Theorems 2,3, and 4 list the most important subgroups. We hope that these theorems will be a useful reference for string theorists. 
\ss
The most surprising finding of this investigation is no doubt the fact that Semispin groups do not contain smaller Semispin groups. This implies that the `` common $SO(16) \times SO(16)$ subgroup of $E_{8} \times E_{8}$ and $Semispin(32)$" simply does not exist, which is obviously a problem for duality. This problem can be overcome by going to a common double cover, but only if certain topological obstructions vanish. In some circumstances, therefore, duality can be obstructed topologically. We shall study this phenomenon elsewhere. 

\vfill\eject
\nt{\bf References}
\ss
\item{[1]}  L. O'Raifeartaigh, {\it Group Structure of Gauge Theories}
            ( Cambridge University Press, Cambridge, 1987 ).	
\ss
\item{[2]}	M.B. Green, J.H. Schwarz, and E. Witten, {\it Superstring Theory }
            ( Cambridge University Press, Cambridge, 1987 ).	
\ss
\item{[3]} J. Kiskis, Phys. Rev. {\bf D17}, 3196 (1978). 
\ss
\item{[4]}	J. Preskill and L.M. Krauss, Nucl. Phys. {\bf B341}, 50 (1990).
\ss
\item{[5]}	B.McInnes, Class. Quantum Grav. {\bf 14}, 2527 (1997).
\ss
\item{[6]}	B.McInnes, J.Phys.A:Math Gen. {\bf 31}, 3607 (1998).
\ss
\item{[7]}	D.J. Gross, J.A. Harvey, E. Martinec, and R. Rohm, Nucl. Phys. {\bf B256}, 253 (1985).
\ss
\item{[8]}	G.G. Ross, {\it Grand Unified Theories} ( Addison - Wesley, Reading, 1984 ). 
\ss
\item{[9]}	M. Berkooz, R.G.Leigh, J. Polchinski, J.H. Schwarz, N. Seiberg, and E. Witten, Nucl. 
            Phys. {\bf B475}, 115 (1996).
 \ss
\item{[10]}	 E. Witten, J. High En. Phys. {\bf 9802-006} (1998). 
\ss
\item{[11]}  J.H. Schwarz, Nucl.. Phys. Proc. Suppl. {\bf 55B}, 1 (1997).
\ss
\item{[12]}  J. de Siebenthal, Comment. Math. Helv. {\bf 31}, 41 (1956).
\ss
\item{[13]} B.McInnes, J. Math. Phys. {\bf 38}, 4354 (1997). 
\ss
\item{[14]}	 H.B. Lawson and M.L. Michelsohn, {\it Spin Geometry} 
             (Princeton University Press, Princeton, 1989).
\ss
\item{[15]}	I.R. Porteous, {\it Clifford Algebras and the Classical Groups} 
            ( Cambridge University Press, Cambridge, 1995 ). 
\ss
\item{[16]}	T. Br{\" o}cker and T. tom Dieck, {\it Representations of Compact Lie Groups } \
( Springer - Verlag, Berlin, 1985 ).
\ss
\item{[17]}	E. Witten,  {\it D-Branes and K-Theory } (hep-th/9810188).
\ss
\item{[18]}	L. Alvarez - Gaum{\'e}, P. Ginsparg, G. Moore, and C. Vafa, Phys. Lett. {\bf B171}, 155 ( 1986 ). 
\ss
\item{[19]}	L. J. Dixon and J.A. Harvey, Nucl. Phys. {\bf B274}, 93 (1986).
\ss
\item{[20]}	J.D. Blum and K.R. Dienes, Nucl. Phys. {\bf B516}, 83 (1998).
\ss
\item{[21]}	P.S. Aspinwall, Nucl. Phys. {\bf B496}, 149 (1997).
\ss
\item{[22]}	J.A. Wolf, Spaces of Constant Curvature ( Publish or Perish, Wilmington, 1984 ).

\bye